\documentclass[oneside,11pt]{article}

\usepackage{graphicx}
\usepackage{changepage}
\newenvironment{myenv1}{\begin{adjustwidth}{-0.2\textwidth}{-0.2\textwidth}}{\end{adjustwidth}}

\usepackage{array}

\newcolumntype{x}[1]{%
 >{\centering\hspace{0pt}}p{#1}}%

\begin{document}
\title{{\Large IRXCT: Iterative Reconstruction and visualization application for X-ray Computed Tomography}}
\author{D. Trinca\\Spitalul Clinic Judetean de Urgenta Arad, Romania\\\\R. Madureira\\CITAB, Vila Real, Portugal}
\date{}
\maketitle

\begin{abstract}
This report describes the IRXCT Windows application for reconstruction and visualization of tomography tasks.

Keywords: iterative reconstruction, tomography, visualization
\end{abstract}

\section{Introduction}
In this report, we describe the IRXCT Windows application \cite{dnt2} for reconstruction and visualization in tomography \cite{KakAC2001,BeisterM2012} that includes the recently proposed SbIR algorithm \cite{dnt1}. Besides the SbIR algorithm, the SART algorithm \cite{SART1} is also included. Adding other iterative algorithms to the application is very easy, as it requires only the addition of a new function for each algorithm. The IRXCT application has been designed so that reconstructions and associated calculations could be done by using the programmed interface; so it has been designed so that it is somehow complementary to other applications where the reconstructions and other calculations are usually programmed manually by writing a script that will run the commands.

Three examples of tomography software available for download and regular use are the following: Astra toolbox, CASToR and NiftyRec.
\begin{description}
\item[Astra toolbox] is a MATLAB and Python toolbox \cite{astra-toolbox1,astra-toolbox2} of high-performance GPU primitives for 2D and 3D tomography. It supports 2D parallel and fan beam geometries, and 3D parallel and cone beam. All of them have highly flexible source/detector positioning. A large number of 2D and 3D algorithms are available, including FBP, SIRT, SART, CGLS. The basic forward and backward projection operations are GPU-accelerated, and directly callable from MATLAB and Python to enable building new algorithms. The source code of the ASTRA Toolbox is available on GitHub.
\item[CASToR] is an open-source multi-platform project \cite{castor} for 4D emission (PET and SPECT) and transmission (CT) tomographic reconstruction. This platform is a scalable software providing both basic image reconstruction features for "standard" users and advanced tools for specialists in the reconstruction field, to develop, incorporate and assess their own methods in image reconstruction (such as specific projectors, optimization algorithms, dynamic data modeling, etc) through the implementation of new classes.
\item[NiftyRec] is a software for tomographic reconstruction \cite{niftyrec}, providing GPU accelerated reconstruction tools for emission and transmission computed tomography.
\end{description}

The IRXCT application has available a number of functions that are complementary to other tomography software like the three software packages described above. For example, one important function is the possibility to see in real-time, that is after each iteration, the current reconstruction along with a convergence plot showing the situations of all iterations that have been already run. Next, we describe in detail all the functionalities currently available in the IRXCT application.

\begin{figure}
\begin{myenv1}
\centering
\includegraphics[width=1.2\textwidth]{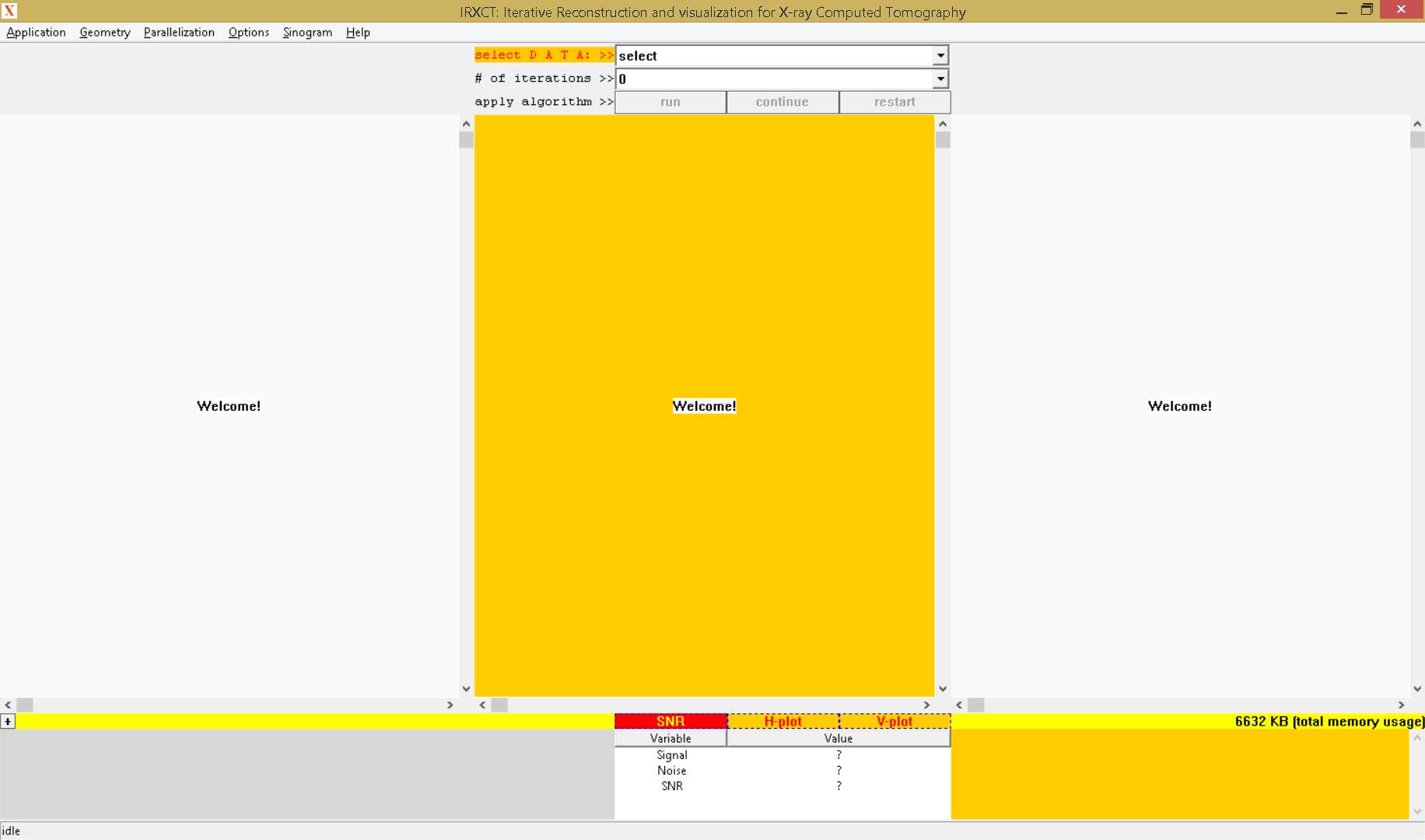}
\caption{\label{fig:01}The IRXCT application at start}
\end{myenv1}
\end{figure}
\section{Description of the IRXCT Application}
The IRXCT application is written in Visual C++ 2015, using only Win32 API functions. Applications that use only Win32 API functions are in general very fast, and this was the main reason for choosing to write all the functionalities using only Win32 API functions. It consists of
\begin{enumerate}
\item a toolbar positioned at the top of the client area;
\item three child windows where the sinogram, the reconstruction (the middle child window) and the original image (for the simulation mode only) are shown;
\item a convergence window in the bottom-left corner where real-time convergence of the reconstruction can be observed;
\item a list view at the bottom where mean signal, mean noise, and SNR (signal-to-noise ratio) calculations are shown;
\item a report window that, during the reconstruction informs the user with the iteration number that is currently under calculation, and at the end of the reconstruction shows a report that informs about the time taken to obtain the reconstruction.
\end{enumerate}

The menu of the application consists of six submenus:
\begin{enumerate}
\item the "Application" submenu has two three items: "Algorithm" which the user can use for choosing the iterative algorithm to be applied (currently there are two choices, SART and SbIR); "Simulation Mode" which is an option that can be checked or not depending on whether the reconstruction is to be done for real sinogram (this corresponds to "Simulation Mode" unchecked, which is the default when starting the application) or for simulated sinogram (this corresponds to "Simulation Mode" checked); "Exit" option which the user can choose to exit the application at any point in time (including during the reconstruction process).
\item the "Geometry" submenu has one item called "Specify Geometry" which the user can user to modify different parameters of the geometry used, including the number of detectors, the number of views, the number of rows and columns of the reconstruction, the distance from source to object (STO) and the distance from source to the detectors' line (STD); when starting the application, these parameters are initially set as nd=512, nv=64, nx=512, ny=512, STO=1024.0, STD=1024.0.
\begin{figure}[t!]
\begin{myenv1}
\centering
\includegraphics[width=1.2\textwidth]{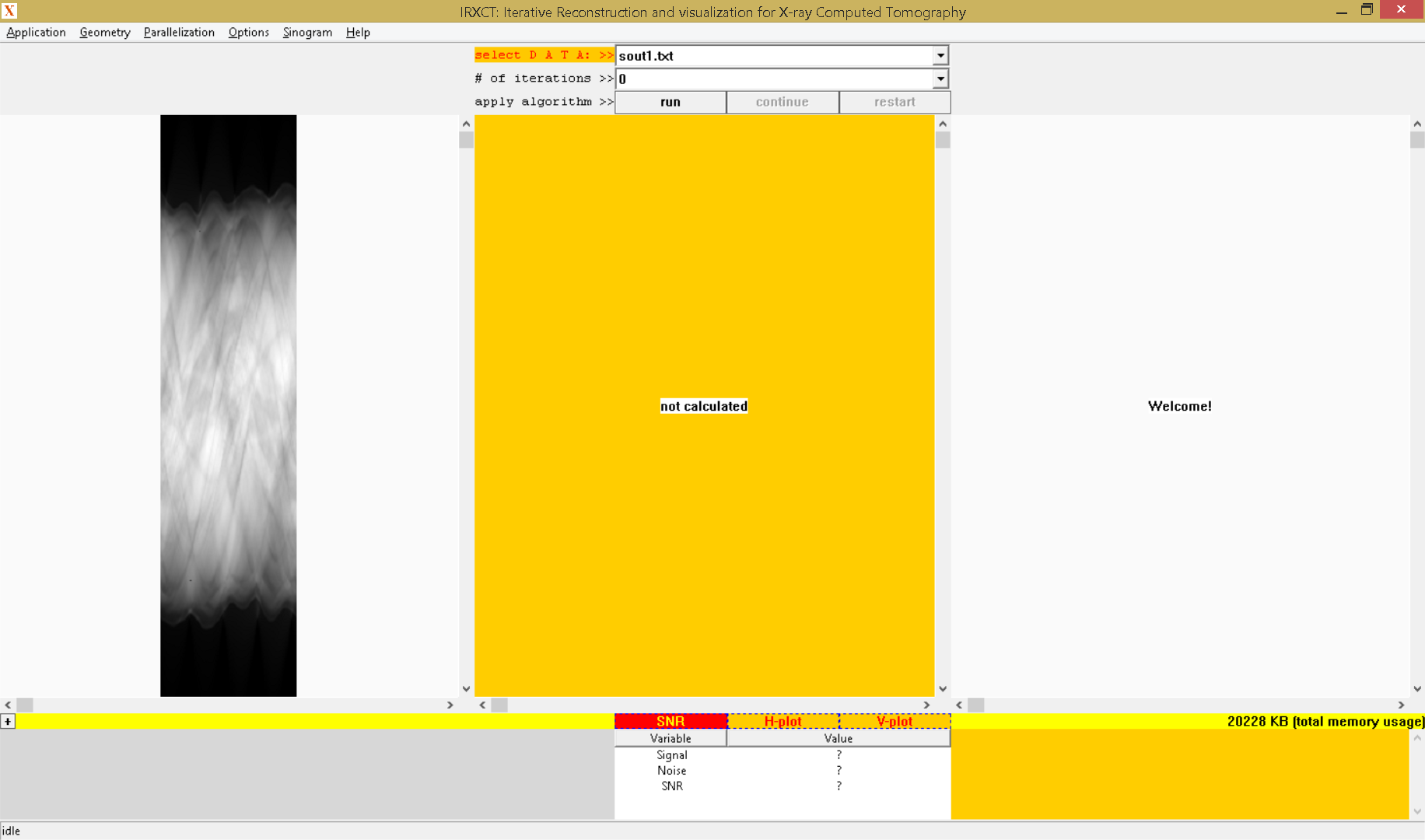}
\caption{\label{fig:02}Selection and displaying of the sinogram}
\end{myenv1}
\end{figure}
\item the "Parallelization" submenu has N options called "1 Thread(s)", "2 Threads", ..., "N Thread(s)" (at any point, only one out of the N items are checked and the others are unchecked), where N is the number of processors that the application detects at start; when starting the application, the "N Thread(s)" is checked meaning that the reconstruction will be done using N threads that the system will usually spread over all N processors available.
\item the "Options" submenu has three items: "Maintain Aspect Ratio" which refers to displaying the sinogram, the reconstruction and the original image (if any) by maintaining the aspect ratio or by using the entire space available; "Show Real-Time Converegence" refers to showing the convergence of the reconstruction and the reconstruction itself after each iteration; "Calculation with Line Integrals" refers to calculations by line integrals or by area integrals (when running reconstructions for real sinograms, area integral is preferred as this corresponds to the way the sinogram has been obtained); "Run with Highest Priority" refers to setting the highest priority for the application (REALTIME$\_$PRIORITY$\_$CLASS), so when this is checked a slightly faster reconstruction is expected.
\item the "Sinogram" submenu has one item called "Add Noise" which the user can use to add noise to the sinogram (in the simulation mode only) before reconstruction.
\item the default "Help" submenu that shows the current version of the application.
\end{enumerate}

\begin{figure}
\begin{myenv1}
\centering
\includegraphics[width=1.2\textwidth]{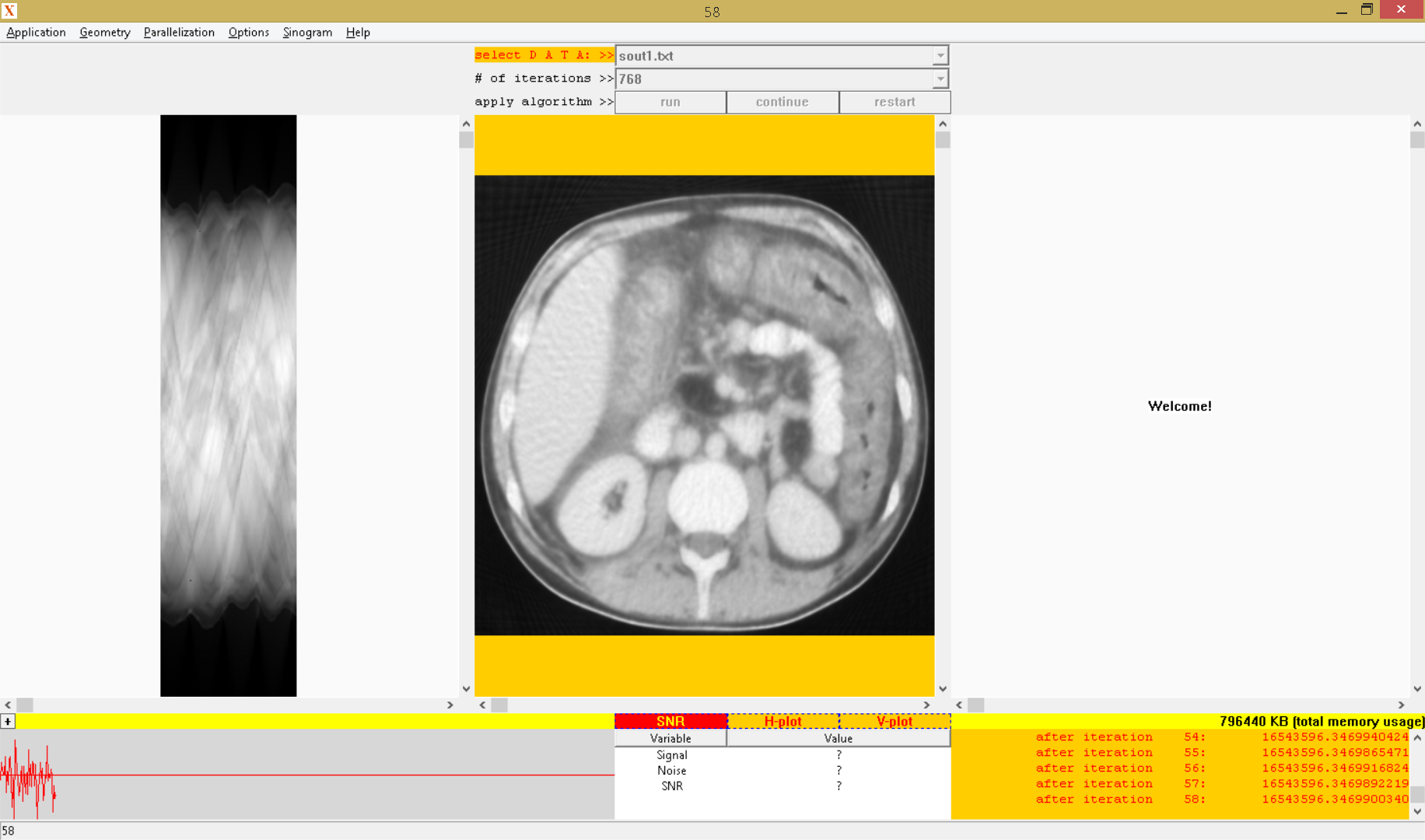}
\caption{\label{fig:03}Situation after iteration 58}
\end{myenv1}
\end{figure}

At start, the application looks like in Figure \ref{fig:01}. The submenu items and their default states (checked or unchecked) at the start of the application are summarized in Table \ref{table:1}.
\begin{table}
\caption{\label{table:1}default states for the menu options of the application}
\centering
\begin{tabular}{|c|c|c|}
\hline
submenu item & checked/unchecked & submenu\\
\hline
\hline
"Algorithm" & n/a & "Application"\\
"Simulation Mode" & unchecked & "Application"\\
"Exit" & n/a & "Application"\\
\hline
"Specify Geometry" & n/a & "Geometry"\\
\hline
"1 Thread(s)" & unchecked & "Parallelization"\\
"2 Thread(s)" & unchecked & "Parallelization"\\
...\\
"N Thread(s)" & checked & "Parallelization"\\
\hline
"Maintain Aspect Ratio" & checked & "Options"\\
"Show Real-Time Convergence" & checked & "Options"\\
"Calculation with Line Integrals" & checked & "Options"\\
"Run with Highest Priority" & unchecked & "Options"\\
\hline
"Add Noise" & n/a & "Sinogram"\\
\hline
\end{tabular}
\end{table}
\subsection{Reconstruction from Real Sinogram}
To show how a reconstruction from a real sinogram is done, we first choose nv=120 (number of views) at the "Geometry" submenu and then choose the file "sout1.txt" sinogram (corresponding to an abdominal cross-section, acquired at Spitalul Clinic Judetean de Urgenta Arad) in the first combobox of the toolbar. This sinogram data consists of 512 $*$ 120 lines, each line containing the real value (double type) of a (detector,view) pair: the first 512 values are for view 1, the next 512 values are for view 2, and so on. After selecting the file "sout1.txt", the sinogram is shown in the left window. Then, we choose the number of iterations (768 in our example here) from the second combobox of the toolbar, and uncheck the item "Calculation with Line Integrals" from the "Options" submenu which means that system matrix and the reconstruction are calculated using area integrals.
Selection of the sinogram and its displaying in the left window is like in Figure \ref{fig:02}. Once a sinogram is selected, the middle window where the reconstruction is to be displayed shows "not calculated".
\begin{figure}[t!]
\begin{myenv1}
\centering
\includegraphics[width=1.2\textwidth]{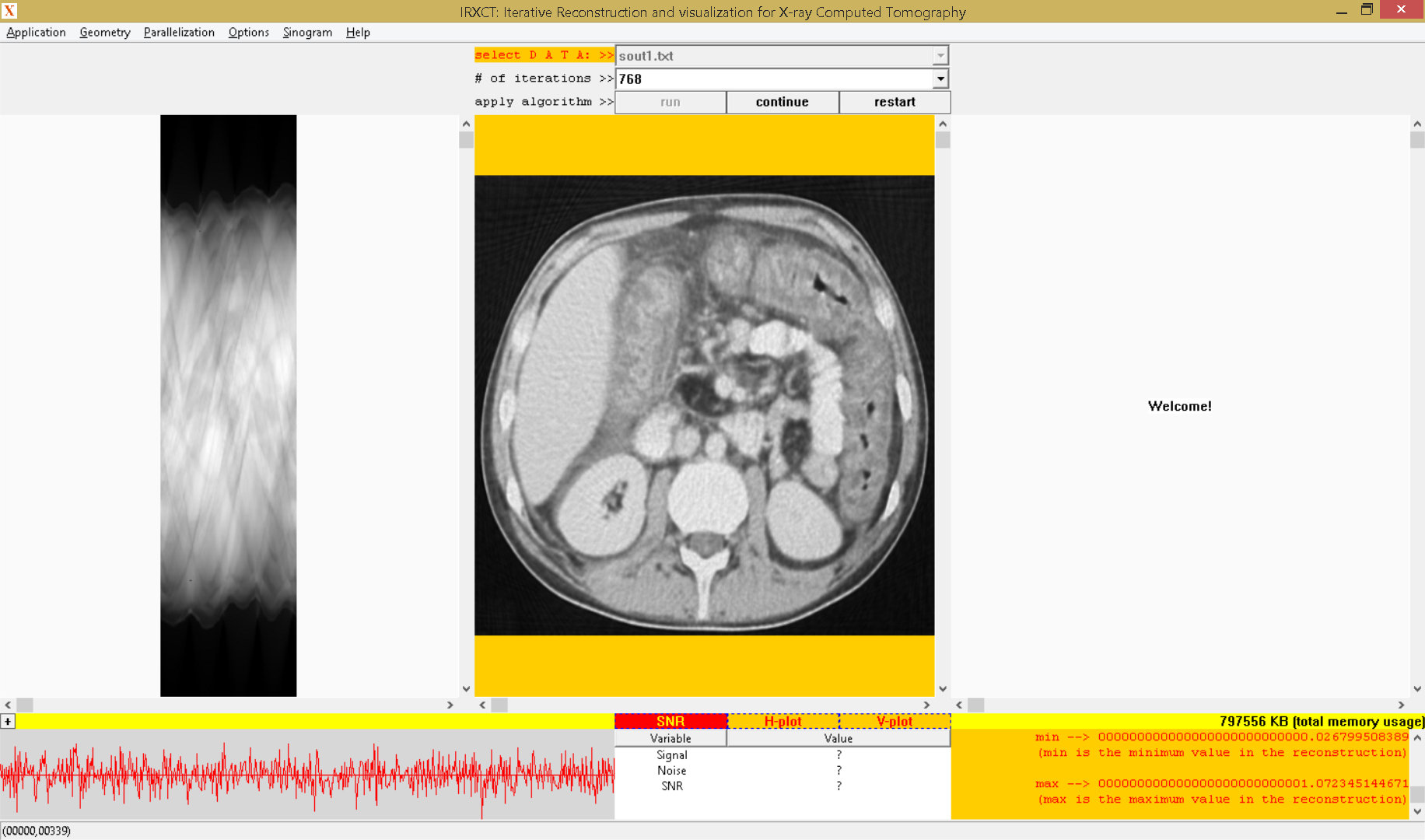}
\caption{\label{fig:05}Situation after running all 768 iterations}
\end{myenv1}
\end{figure}

With the sinogram and the number of iterations chosen, we press the "run" button so that calculation of the system matrix (using area integrals) and then the reconstruction start. During the calculation of the system matrix the yellow static control right above the report window shows in real-time the current memory usage of the application. Once the system matrix is calculated, the iterations start and after each iteration the reconstruction window shows the reconstruction updated and the convergence window shows the plot of the convergence. For example, Figure \ref{fig:03} shows the situation after iteration 58. In Figure \ref{fig:05} it is shown the reconstruction after all 768 iterations are completed. After all 768 iterations are completed, the report window shows a report consisting of four indicators:
\begin{enumerate}
\item "time 1" is the time (double type) taken to calculate the system matrix (in this example, it is 38.006 seconds)
\item "time 2" is the time (double type) taken to run the 768 iterations (in this example, it is 315.007 seconds)
\item "min" is the minimum value (double type) in the reconstruction (in this example, it is approximately 0.026)
\item "max" is the maximum value (double type) in the reconstruction (in this example, it is approximately 1.072)
\end{enumerate}

\begin{figure}
\begin{myenv1}
\centering
\includegraphics[width=1.2\textwidth]{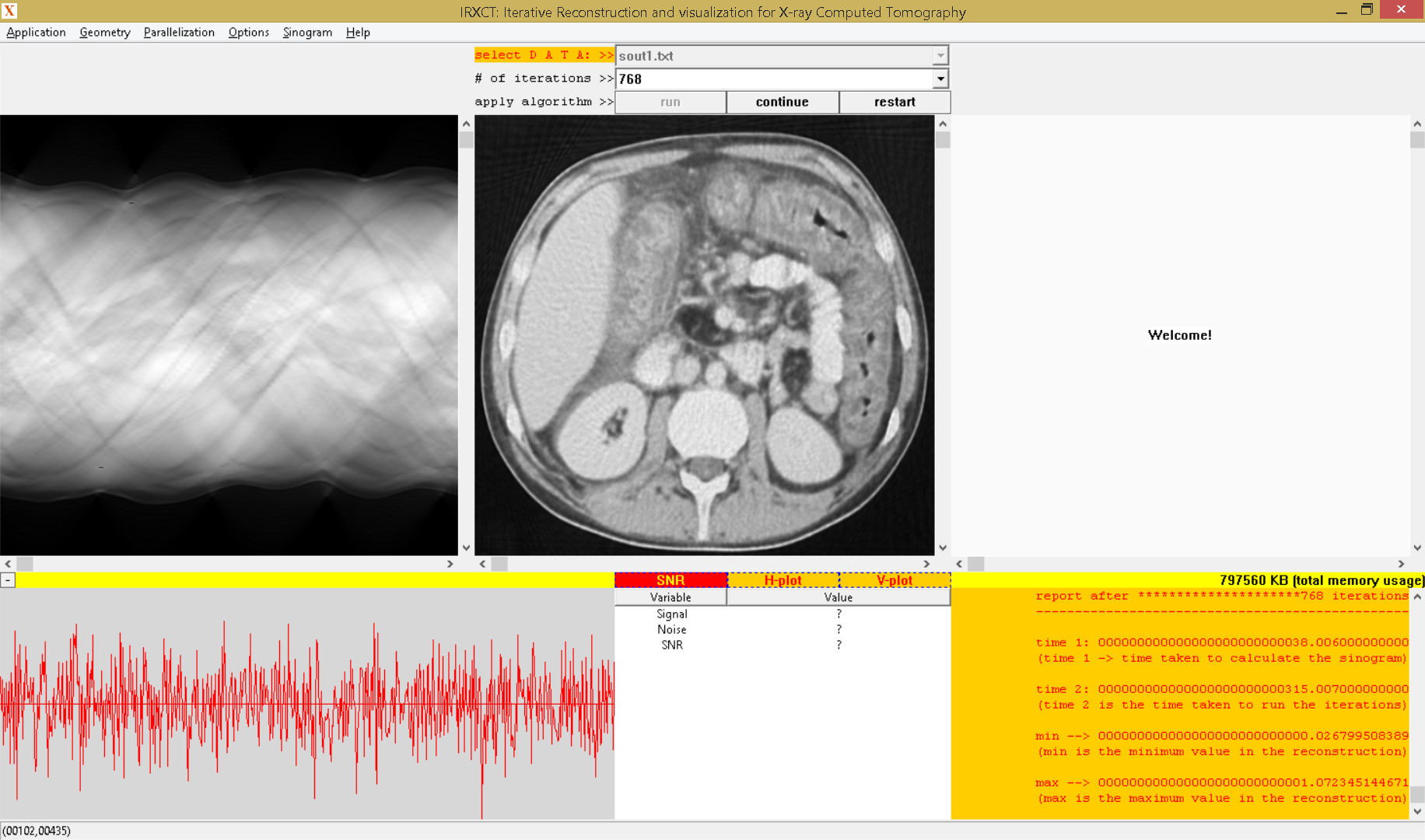}
\caption{\label{fig:06}Increased space for the bottom part of the application}
\end{myenv1}
\end{figure}

At any point, before reconstruction, during reconstruction or after reconstruction, the user can check or uncheck the "Maintain Aspect Ratio" item of the "Options" submenu so that each image is displayed by using the entire space available or by maintaining aspect ratio. Also, the small "+" button right above the convergence window can be used to increase the space used by the bottom part of the application from $15\%$ to $35\%$ as in Figure \ref{fig:06}.

After all iterations are completed and the final reconstruction is shown, the "continue" button can be used to continue running another set of iterations, starting with the currently obtained reconstruction. The "restart" button restarts everything and cleans up the used memory space so that the user can check, choose, or uncheck different options and then run iterations starting again from the initial solution. Also, once the final reconstruction is shown, the user can move with the mouse pointer over the reconstruction window and calculate either mean signal for a rectangle (chosen with left mouse button) or mean noise for a rectangle (chosen with the right mouse button). When both signal rectangle and noise rectangle are chosen, the SNR is automatically calculated and shown in the list view control at the bottom of the application. For choosing a different signal or noise rectangle the user just needs to choose the desired new rectangle, as the current rectangle will be automatically deselected. A currently selected rectangle can also be manually deselected by clicking with the respective mouse button (left button for signal, or right button for noise) anywhere on the reconstruction.

\begin{figure}
\begin{myenv1}
\centering
\includegraphics[width=1.2\textwidth]{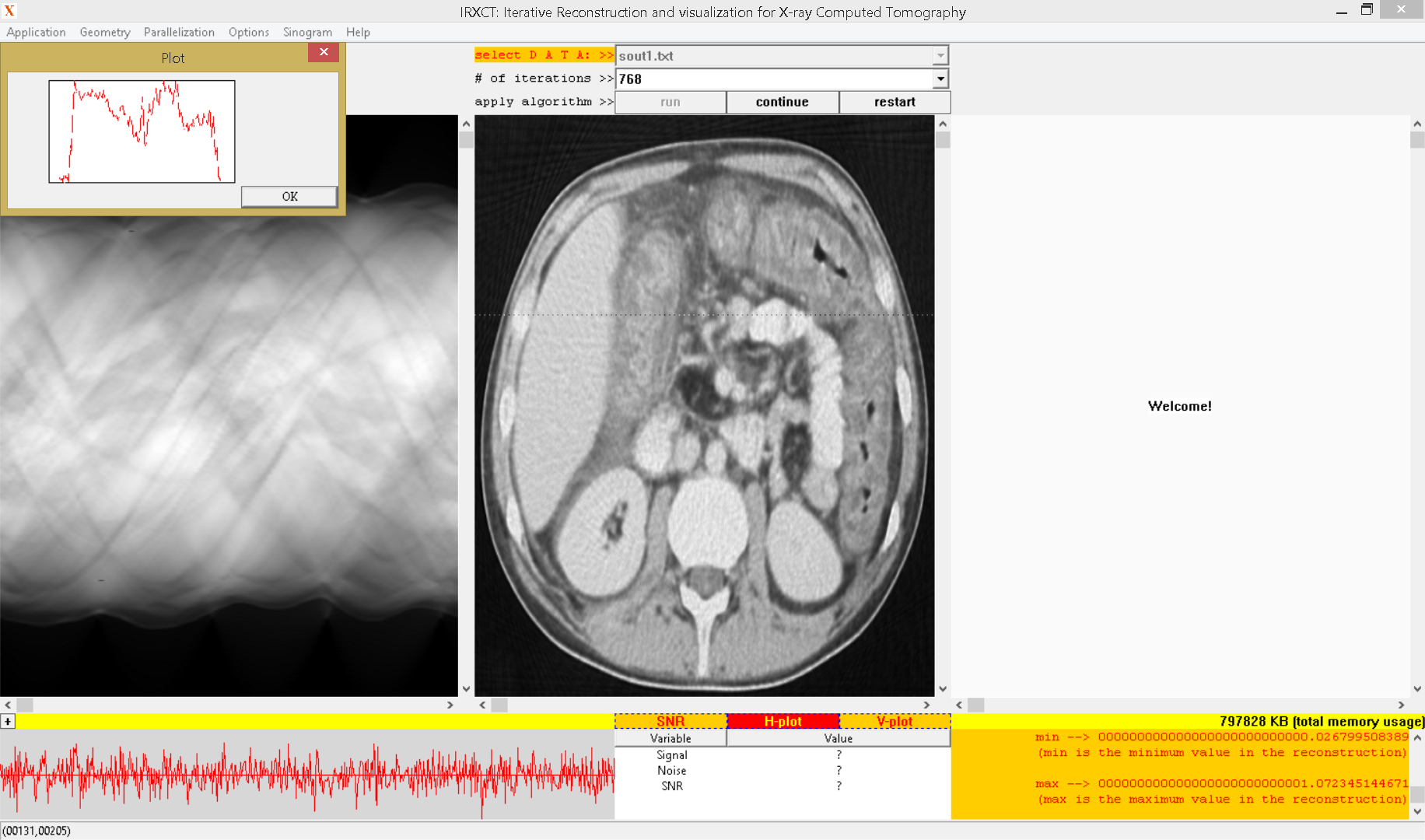}
\caption{\label{fig:08}Plot showing a line (highlighted with a dotted line in the reconstruction window) of the reconstruction, when the "H-plot" button is checked}
\end{myenv1}
\end{figure}

If the "H-plot" button is checked then the user can click on any (line l, column c) point of the reconstruction and a plot showing line l will be displayed (in a separate dialog box, like in Figure \ref{fig:08}).

If the "V-plot" button is checked then the user can click on any (line l, column c) point of the reconstruction and a plot showing column c will be displayed.
\begin{figure}
\begin{myenv1}
\centering
\includegraphics[width=1.2\textwidth]{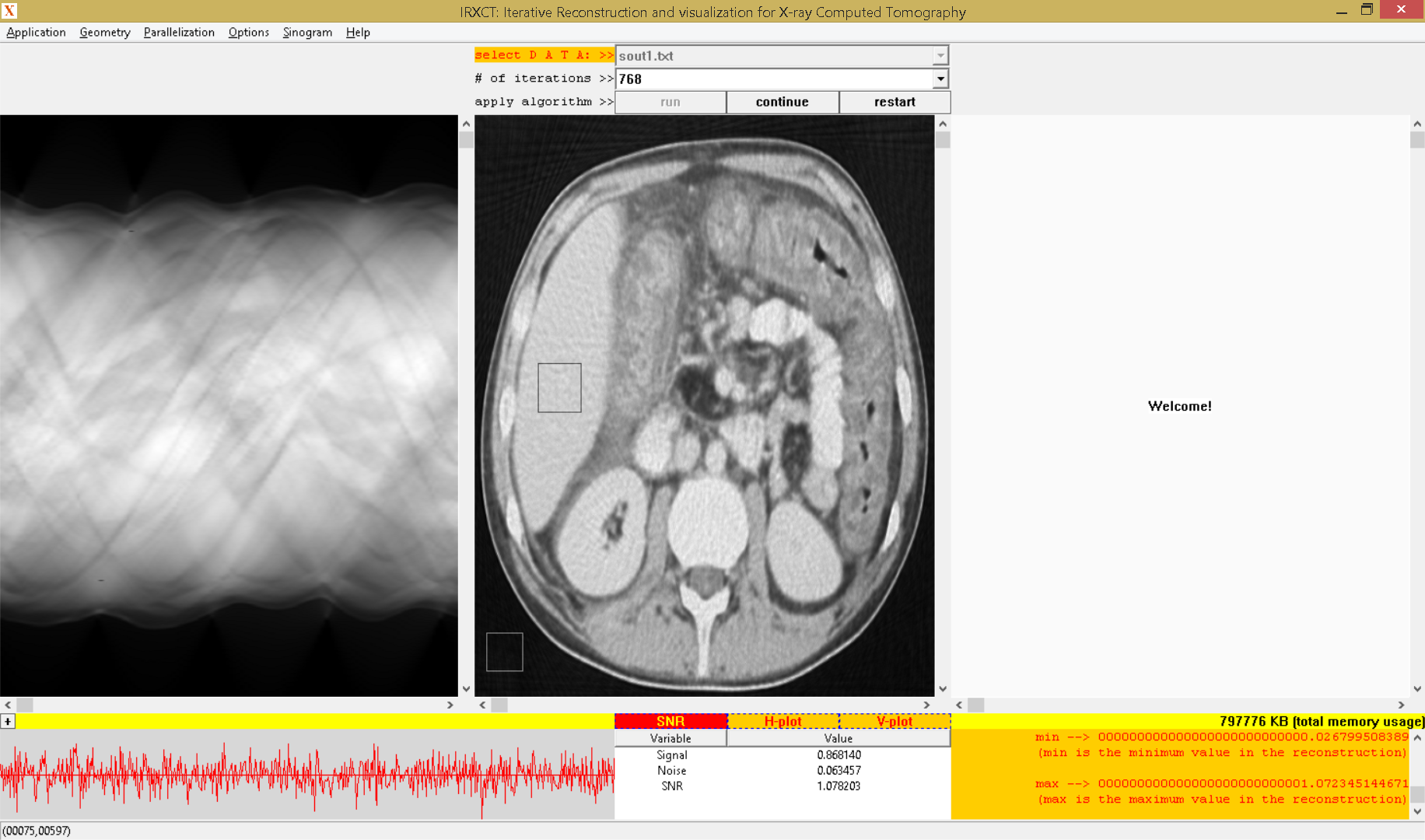}
\caption{\label{fig:07}when the "SNR" button is checked, the user can choose a signal rectangle and also a noise rectangle; when both rectangles are present the SNR value is automatically calculated and shown in the list view control at the bottom of the application}
\end{myenv1}
\end{figure}
\subsection{Reconstruction from Simulated Sinogram}
\begin{figure}
\begin{myenv1}
\centering
\includegraphics[width=1.2\textwidth]{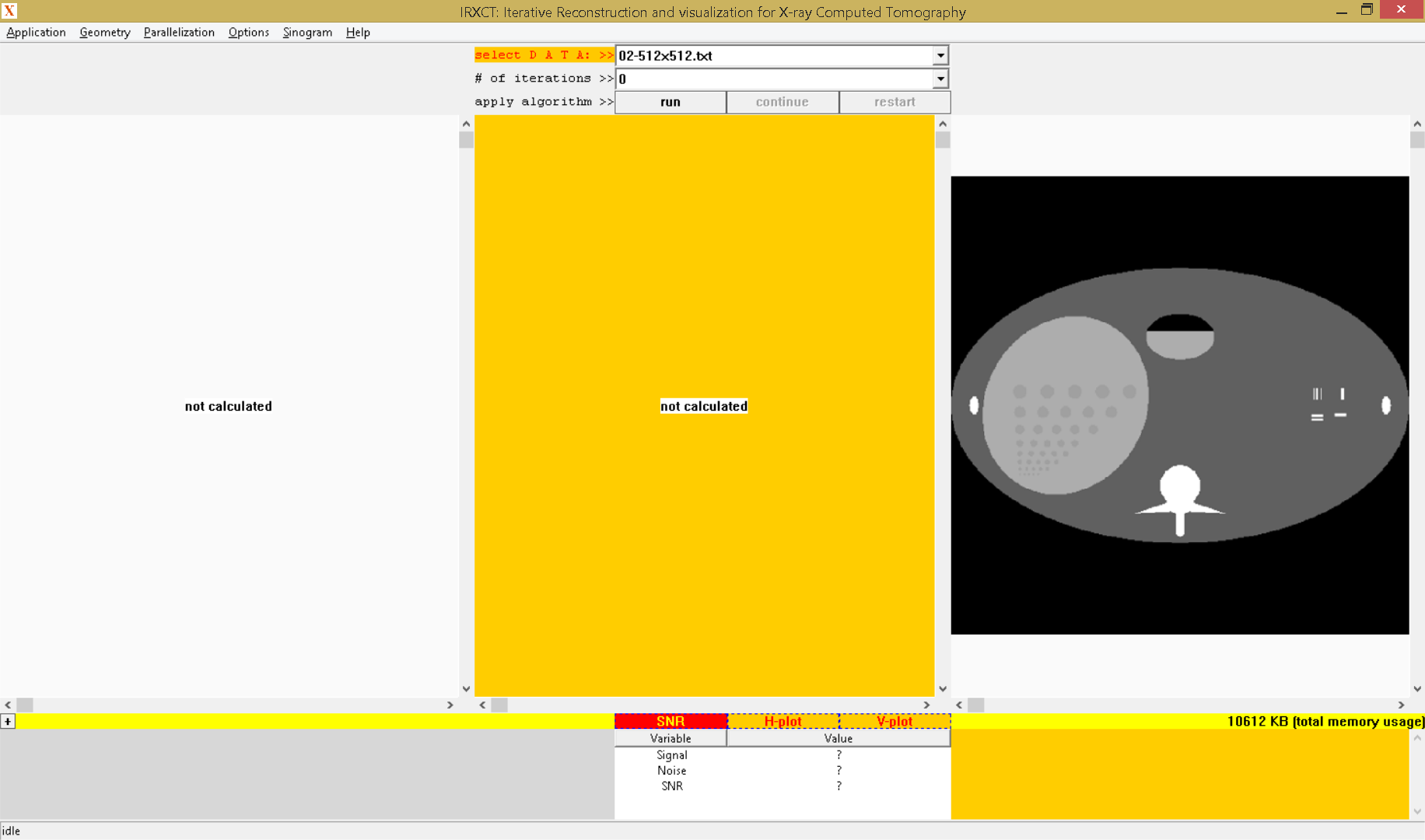}
\caption{\label{fig:21}Abdominal phantom, of size 512 by 512 pixels}
\end{myenv1}
\end{figure}

For showing reconstruction from a simulated sinogram (which means the "Simulation Mode" menu item is checked), we choose the abdominal image \cite{pd1} shown in Figure \ref{fig:21} by selecting the file "02-512x512.txt" in the data combobox of the toolbar. After setting the number of iterations to 768 again, and also making sure that the "Calculation with Line Integrals" is unchecked, and making sure that nv=120 again, we press the "run" button so that sinogram is calculated (once sinogram is calculated, it is shown in the left window) and then the 768 iterations are run.

\begin{figure}[t!]
\begin{myenv1}
\centering
\includegraphics[width=1.2\textwidth]{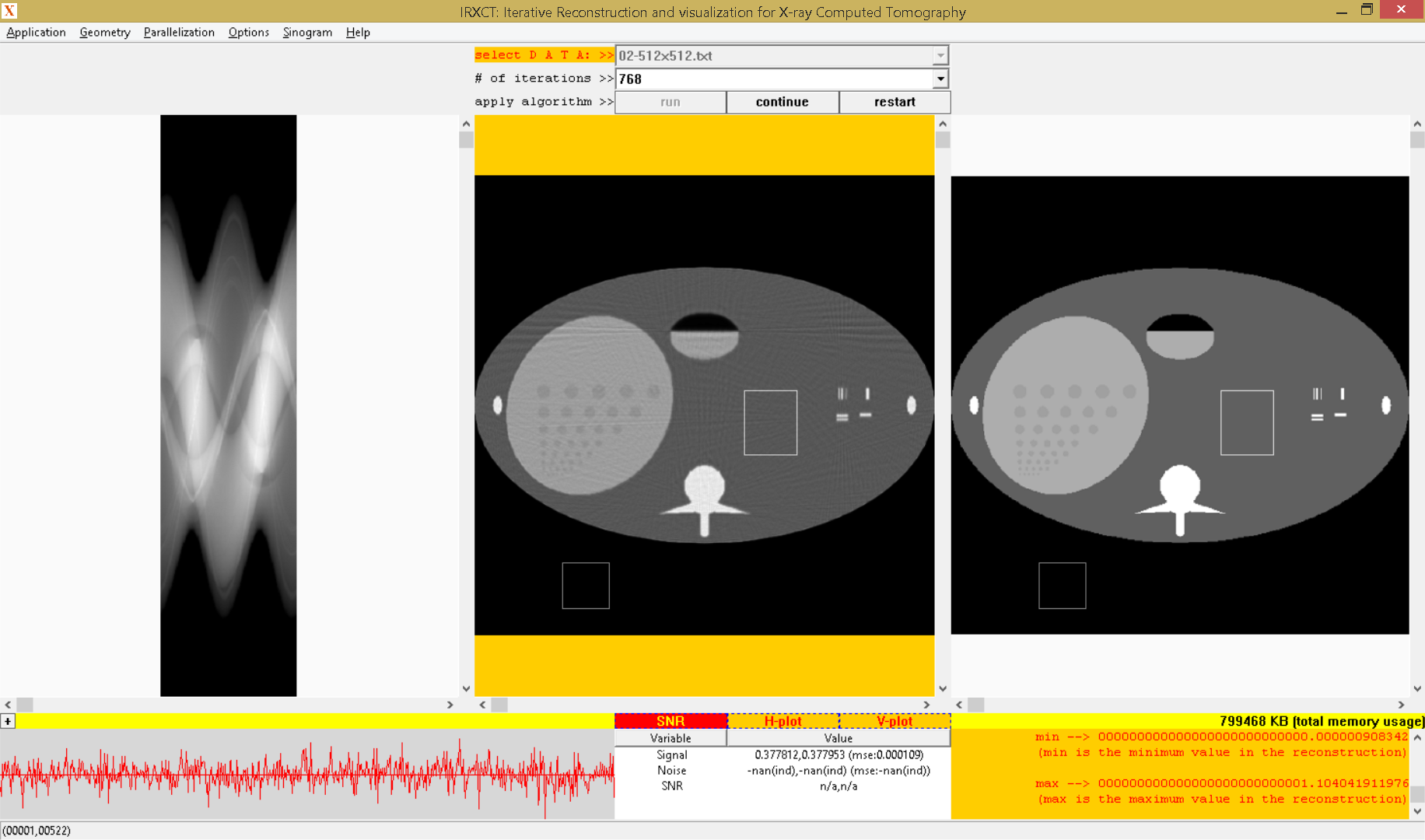}
\caption{\label{fig:23}Situation after all 768 iterations are completed}
\end{myenv1}
\end{figure}

Once the iterations are started, the reconstruction window shows the updated reconstruction and the convergence window shows a real-time plot of the convergence. The situation after all 768 iterations are completed is shown in Figure \ref{fig:23}.

Again we can choose signal and noise rectangles, but if noise for example is 0 then SNR will not be available ("n/a"). If SNR is very large ($>$ 10000.0) then SNR will show "$>$ 10000.0". The SNR is calculated using the formula
\begin{displaymath}
\frac{\textnormal{mean of the signal region}}{\textnormal{standard deviation of the noise region}}.
\end{displaymath}

In the simulated mode, when choosing a rectangle in the reconstruction window, then the same rectangle is chosen in the original image, and values are calculated for both images (first value for reconstruction, second value for original image). In this example, for the signal rectangle shown in Figure \ref{fig:23}, the first value for mean signal is 0.377812 (for the reconstruction) and the second value for mean signal is 0.377953 (for the original image). In the simulated mode, mean-square-error (mse) of the two regions is also calculated automatically and shown as "mse:value".

\begin{figure}
\begin{myenv1}
\centering
\includegraphics[width=1.2\textwidth]{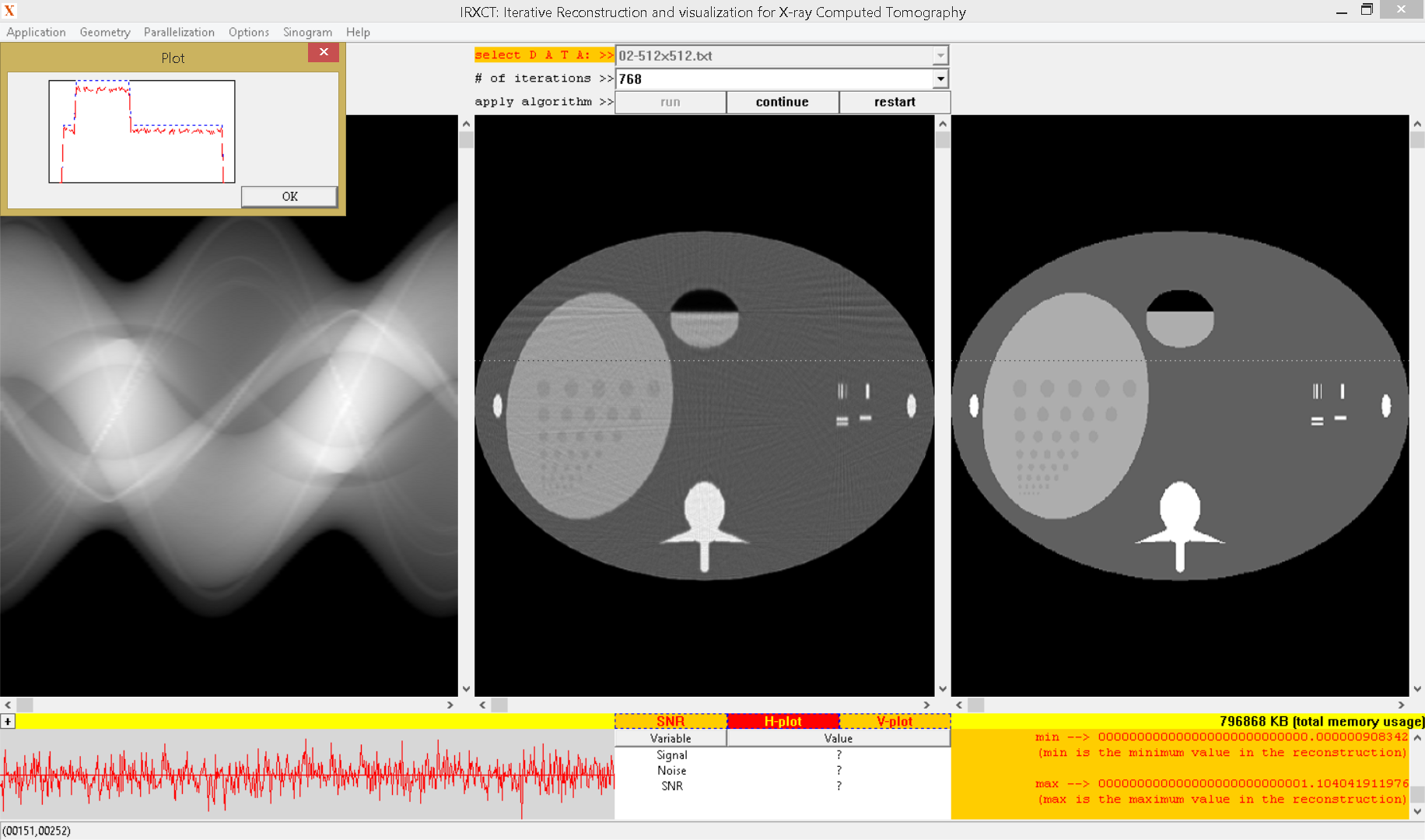}
\caption{\label{fig:24}Plot showing the selected line (in blue for the original image, and in red for the reconstruction)}
\end{myenv1}
\end{figure}

Also, since we are now in simulated mode, choosing a line of the reconstruction when "H-plot" is checked, will show a comparison plot between the two lines (in blue for the original image, and in red for reconstruction), as in Figure \ref{fig:24}.
\subsection{Region-of-interest Tomography}
In the simulated mode, before the user presses the "run" button a region-of-interest (roi) of the original image can be selected using the left mouse button. In such case, the reconstruction will be done only for the selected roi.
\subsection{Running with Highest Priority}
When the "Run with Highest Priority" option in the "Options" submenu is checked, the priority class is changed to the highest possible. For a reconstruction that normally takes a few seconds the improvement may not be so visible, but for longer reconstructions this would decrease the execution time with a couple of seconds or so.

\end{document}